\documentclass{IEEEcsmag}

\usepackage[colorlinks,urlcolor=blue,linkcolor=blue,citecolor=blue]{hyperref}

\usepackage{upmath}
\usepackage[utf8]{inputenc}
\usepackage[T1]{fontenc}
\usepackage{balance}

\jvol{XX}
\jnum{XX}
\paper{8}
\jmonth{May/June}
\jname{IEEE Computer Graphics and Applications}
\pubyear{2021}

\begin{document}

\sptitle{Department: People in Practice}
\editor{Editors: Melanie Tory, m.tory@northeastern.edu \\Daniel Keefe, dfk@umn.edu}

\title{Communicating Patient Health Data: A Wicked Problem}

\author{Fateme Rajabiyazdi}
\affil{Carleton University}

\author{Charles Perin}
\affil{University of Victoria}

\author{{L}ora Oehlberg}
\affil{University of Calgary}

\author{Sheelagh Carpendale}
\affil{Simon Fraser University}

\markboth{People in Practice}{Communicating Patient Self-Collected Health Data in the Clinic: A Wicked Problem}

\begin{abstract}
Designing patient-collected health data visualizations to support discussing patient data during clinical visits 
is a challenging problem
due to the heterogeneity of the parties involved: patients, healthcare providers, and healthcare systems.
Designers must  
ensure that all parties' needs are met.
This complexity makes it challenging
to find a definitive solution that can work for every individual.
We have approached this research problem -- communicating patient data during clinical visits -- as a wicked problem. 
In this article, we outline how wicked problem characteristics 
apply to our research problem. We then describe the research methodologies we employed
to explore the design space of individualized patient data visualization solutions. Last, we reflect on the insights and experiences we
gained through this exploratory design process. We conclude with a call to action for 
researchers and visualization designers to consider patients' and healthcare providers'
individualities when designing patient data visualizations.
\end{abstract}

\maketitle

\chapterinitial{Designing visualizations} to represent patient self-collected health data is a challenging problem.   
With the growing number of self-monitoring technologies, many patients monitor their health at home and record their data~\cite{Choe2021}.
Patients often bring these self-collected data to clinical visits and share them with healthcare providers.
The complexity of this data makes it difficult for healthcare providers to review the data,
make clinical judgments, and provide tailored suggestions  in short clinic visits. A potential solution to 
accurately show a summary of patient-collected health data is to visualize them.

Researchers and designers need to consider all parties involved in this communication dynamic - patients, healthcare providers, and healthcare systems - when designing visualizations.
This complexity makes it challenging for designers
to find a definitive, general design solution that works for any individual in any situation.

We have been closely collaborating with patients, healthcare providers, and health researchers 
from a local hospital in the province of Alberta, Canada for 5 years to design patient health data visualizations that facilitate clinical communication.
Throughout this multi-year collaboration, 
we came to the conclusion that our research problem is a \textit{wicked problem} (a detailed definition of a wicked problem in provided in the next section) and we cannot solve our research problem using a generalized design solution. 
Thus, to address this research question we
employ a series of exploratory methodologies and propose various individualized designs. 

We take qualitative approaches (interviews and focus groups) to
better understand this wicked problem from patients’ and providers’ perspectives.
From this understanding, we employ an iterative design cycle approach, exploring specific visualization designs developed for individual patients and their healthcare provider, based on patients' unique needs and challenges. 

In this article, we reflect on our experiences in designing and implementing 
individualized patient health data visualizations through this exploratory design process. 
We identify the challenges of designing visualizations
for patients and healthcare providers and discuss future directions. 

The results of all our empirical studies revealed the importance of designing visualizations by considering individuals.
We believe our approach to addressing this challenging real-world wicked problem and the insights we gained through our empirical studies will impact future design studies in the visualization field. 

\begin{figure*}[h!]
    \centerline{\includegraphics[width=\textwidth]{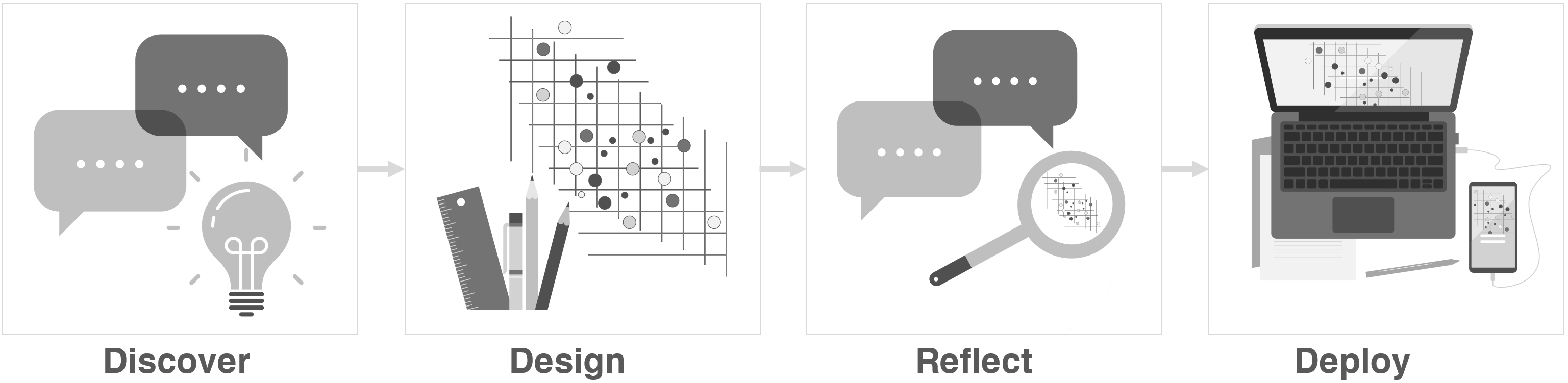}}
    \caption{Iterative exploratory process, Patient-Center Design, to design patient health data
    visualization. We started this process by conducting interviews and focus groups with healthcare providers and 
    patients to \textit{`Discover'} challenges they face reviewing health data. Next, we began to \textit{`Design'} potential 
    visualizations to present patient health data. Then, we asked healthcare providers to \textit{`Reflect'} on the visualization designs. Upon analyzing the reflections, we collaborated with professional developers to \textit{`Deploy'} the designs.}
    \label{fig:method}
\end{figure*}

\section{WICKED PROBLEM CHARACTERISTICS}
The term ``wicked problem'' was originally used in social planning~\cite{Rittel1973}. 
Wicked problems are problems with a high level of conflicts (between stakeholders) on the definition of the problem and its solution.  
 There are many examples of wicked problems such as climate change and obesity.

Our research problem, communicating patient-collected health data, is a \textit{wicked problem}. Many stakeholders are involved in this problem,
including patients, patient family members, healthcare provider teams, and healthcare systems. Each stakeholder defines the problem from their perspective. For example, patients may be most worried about how to share their data effectively. 
The healthcare provider team may think the problem is to find the time in the clinic to review the data. The healthcare system technicians may be concerned with the privacy and security of storing this data.
This complexity of the problem makes it challenging to find a definitive visualization solution to present patient-collected health data that can work for all the stakeholders. 
Here, we discuss in detail the wicked problem characteristics~\cite{Rittel1973} that specifically apply to our research problem.

\textbf{Wicked problems have no definitive formulation.} 
Understanding a wicked problem and anticipating all the possible solutions are tightly related.
We need knowledge of all possible solutions to understand a wicked problem.

With all stakeholders (patients, providers, and healthcare systems) involved in communicating patient-collected data, it is not possible to fully define the problem 
with a pre-set list of specifications. Thus, it is challenging to design one visualization that meets all stakeholders' requirements without complete knowledge of the problem.

\textbf{Wicked problem solutions cannot be black or white.}
Since there is no definitive formulation of a wicked problem, 
it is not easily possible to find definitive solutions, and there is always room for improvement. Thus, solutions to a wicked problem cannot be fully white or black. 

Similarly, we cannot consider patient health data visualizations as definitive right or wrong solutions. 
Since stakeholders with different goals and perspectives are involved in this problem, finding a perfect solution for all parties is not possible. Instead, each stakeholder can judge the proposed visualizations as better, worse, or good.

\textbf{Wicked problems have no ultimate solution tests.} Even though the proposed wicked problem solutions target a group of people, these solutions can affect an extended group of people over time.
Thus, researchers cannot evaluate solutions proposed for a wicked problem completely, as there is no ultimate test of a solution.

It is extremely difficult to evaluate all advantages and disadvantages of the proposed patient health data visualizations.
Using these visualizations may generate consequences on patients' health outcomes, their caregivers' lives, and the dynamic of healthcare provider practices.

\textbf{Wicked problems have no exhaustive set of potential solutions.}
Since there is no definitive set of criteria, researchers cannot claim that they identified all potential solutions. It will be up to stakeholders to 
choose a solution for implementation based on their budget, allocated time for the project, and available resources.

Designing and developing all possible variations of patient health data visualization is not feasible.
Rather, it is a matter of choosing a solution by taking resources and involved stakeholders' perspectives into consideration.

\textbf{Solutions to wicked problems are a one-shot operation.} Every proposed solution is a one-shot operation since every attempt has irreversible consequences. The impact of implementing each solution cannot be undone. 

Similarly, using proposed visualizations by patients, patient family or caregivers, or healthcare providers can have non-reversible influences on their lives or practices.

\textbf{Wicked problems can be symptoms of another wicked problem.}
So, proposing solutions to each sub-problem (symptom)
with the hope to aggregate these solutions to address the wicked problem as a whole is not entirely possible.

Designing patient health visualizations to facilitate communicating patient data can impact the communication dynamics in the clinic, which is another wicked problem.
Thus, designing ad-hoc visualization solutions cannot simply lead us to solve this problem as a whole.

\textbf{The choice of a wicked problem's explanation can determine the nature of the problem’s solution.} 
Each stakeholder involved may provide an explanation from their perspective
that can change the focus of the solution. Due to the unique properties of a wicked problem and the inability to rigorously evaluate a solution, there is no way to test or find the ``correct'' explanation.

\begin{figure*}
    \centerline{\includegraphics[width=\textwidth]{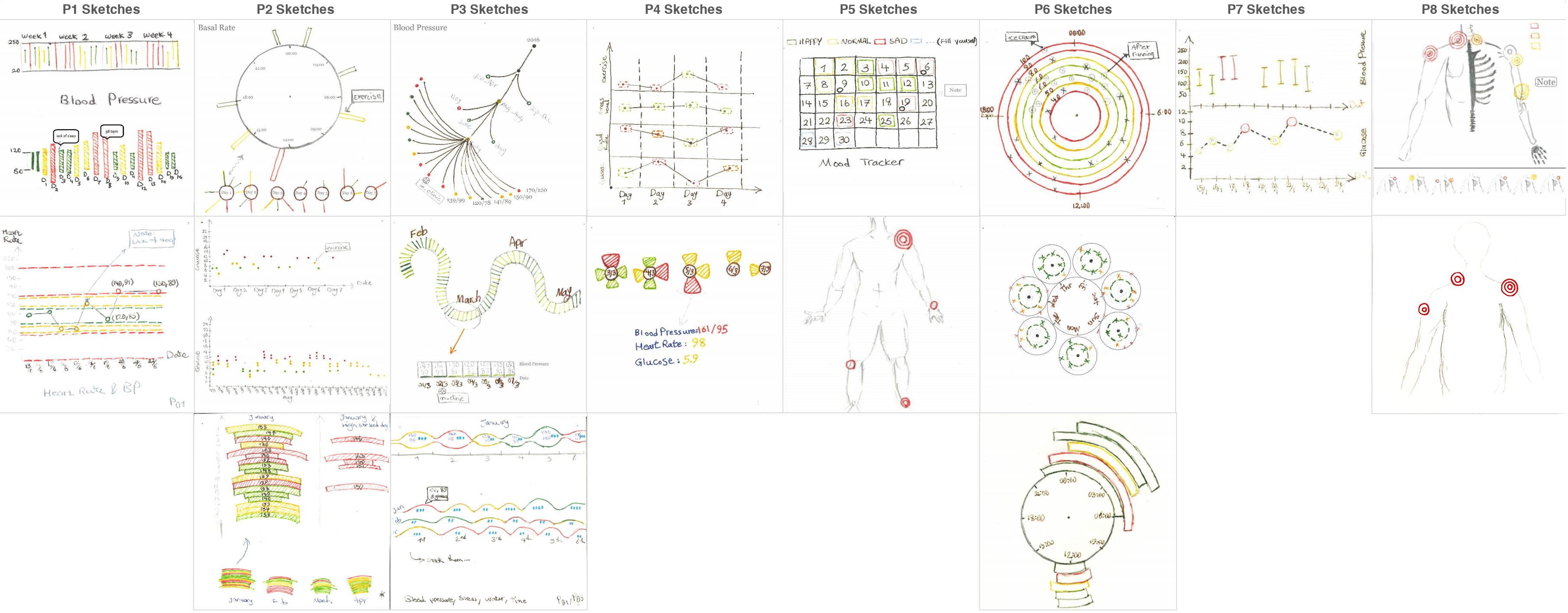}}
    \caption{This design board represents our proposed patient health data visualizations. Each column in this board represents various designs for one patient. We included alternative visualization designs in rows for each patient based on their condition, lifestyle, types of data collected, and relationship with healthcare providers.}
    \label{fig:designBoard}
\end{figure*}

Researchers may associate the problems of communicating patient-collected health data with the way patients organize and present the data and propose visualizations to support patients in presenting their data effectively. 
It is still not possible to thoroughly test if the explanation was the “correct” explanation as many external factors could have affected communication dynamics.



\section{PATIENT-CENTERED DESIGN FOR INDIVIDUALS \& LESSON LEARNED}

There are three main approaches in the literature to address wicked problems: authoritative, competitive, and collaborative~\cite{Roberts2000}. 
To address our research question, communicating patient health data, we took a collaborative approach that 
requires the involvement of all stockholders to find the best solutions and make decisions.

There are recent movements in healthcare for delivering patient-centered care, where the care is respectful and responsive to individual patients' experiences, needs, values, and preferences instead of focusing on the illness~\cite{Barry2012}. Similarly, to design effective technologies or visualizations for healthcare, we need to tailor the design to individual needs, values, and preferences.


Our methodology combines these two approaches: people-centered design and patient-centred care, or as we call it, \textit{patient-centered design}. In this approach, we need to invite patients as partners with healthcare providers in the design process.
Patient-centered design is a cyclic design methodology of discovering the context of use and patient and healthcare provider requirements, designing through fast prototyping, reflecting through getting feedback from patients and healthcare providers, analyzing and refining the design, and developing and integrating the designs into the healthcare systems (See Figure~\ref{fig:method}).

Reviewing the literature and conducting interviews with healthcare providers revealed the most common communication challenges between patients and healthcare providers during clinical visits~\cite{Rajabiyazdi2017}.
After consulting with our healthcare provider collaborators, they pointed out that discussing 
patient self-collected health data is one of the main challenges they face in clinical visits.
\begin{figure*}
    \centerline{\includegraphics[width=\textwidth]{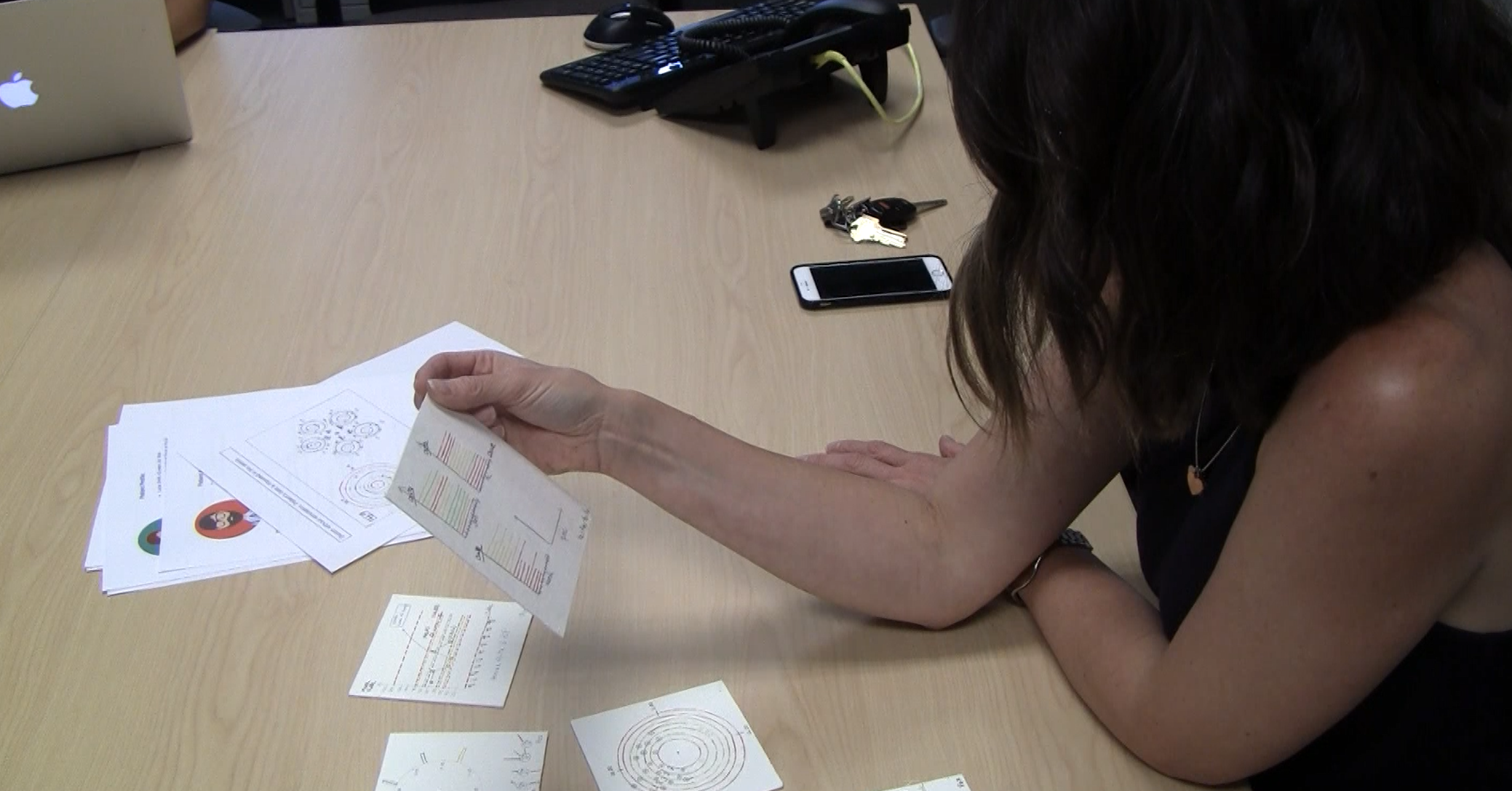}}
    \caption{A healthcare provider reflecting on our proposed visualizations. The results of provider interviews revealed the importance and necessity
   of not only designing customizable visualizations for patients, but also considering the goals of the individual providers when using the visualizations. With all providers, we observed an initial gravitation to familiar visualizations but also a gradual increasing interest in innovative visualizations.}
    \label{fig:HC-FG}
\end{figure*}

\subsection{Design results}
To understand the challenges patients face when collecting, reviewing, and presenting their health data and understanding healthcare providers' needs when requesting and reviewing patient health data, we employed qualitative methodologies.
As with any empirical research method, qualitative methods have benefits and pitfalls~\cite{McGrath1995}.
Using these methods allowed us to gain a rich and real understanding of  
patients and providers. However, we captured the perspectives of 
a small sample size, and our results are not generalizable to all 
patients and providers. 

We conducted a focus group with a mixed group of healthcare providers, including the group of providers who initially approached us and several other providers and researchers who were recruited through word
of mouth in the same hospital. We asked healthcare providers about their experiences reviewing, analyzing and understanding the patient-generated data and giving advice to patients based on their data.

Later, we conducted an hour-long semi-structured interview with eight patients with one or multiple chronic conditions and asked patients to share a sample of their health data collected at home and walk us through the details of the data. After analyzing the interview data, we designed various individually tailored visualizations based on patient stories and the health data they collected at home and shared with us (See Figure~\ref{fig:designBoard}).

\subsection{Implementation results}
There are many methods to evaluate visualization designs. One way is to conduct observational studies in the medical clinics
to monitor and analyze how patients and healthcare providers use and interact with 
data visualizations. However, due to the sensitivity of the
topics discussed during clinical visits, patients and healthcare providers may not be
willing to discuss their routine topics or show hesitations when sharing in the presence of a third party (a researcher). In addition, some challenges
may only appear after lengthy observations of many clinical visits. 
In an interview or focus group study, we have the advantage of asking direct questions relevant to the topic of interest and obtain patients’ and providers’ perspectives over a short time. 

\begin{figure*}
    \centering
    \includegraphics[width=\textwidth]{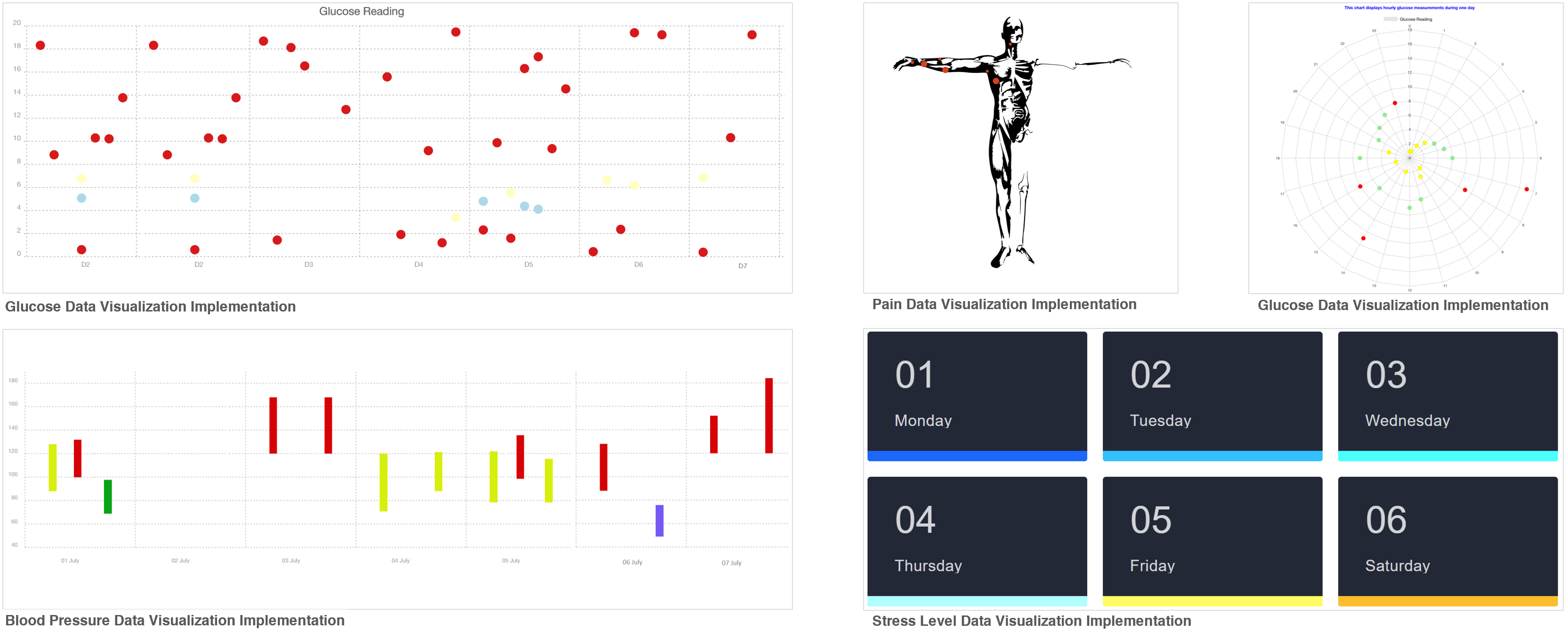}
    \caption{Prototype implementation of the five selected visualization designs after discussions with the healthcare provider team. Each prototype represents one patient data visualization sketch. On the top row (from left), we display blood glucose, pain intensity, and another representation of blood glucose data. On the bottom row, we show blood pressure and stress levels visualizations. In an iterative process, we gained providers' feedback on these designs. For example, providers found the radar glucose data visualization (top left) cluttered for patients, so we removed the crossing lines but also provided two glucose data visualizations.} 
    \label{fig:Design1-2}
\end{figure*}

Thus, we decided to conduct interviews with healthcare providers to evaluate the receptions of our visualization designs
~\cite{Rajabiyazdi2020} (See Figure~\ref{fig:HC-FG}).  
We took our visualization designs back to three providers, who
were among the group that initiated this project, seeking their feedback.
In collaboration with the healthcare providers, we selected six visualization sketches that they saw value in incorporating into the patient-centred care plan.
The developer team found one visualization design (P3 sketch on the first row Figure~\ref{fig:designBoard}) not easy to implement within the set budget, so we removed this design. 
We got feedback iteratively from the healthcare provider and the developer team and adjusted our designs for integration into our Alberta patient care plan system (See Figure~\ref{fig:Design1-2}).
Lastly, we removed the pain tracking design
as the healthcare provider team found it repetitive to other existing applications. 

\section{CHALLENGES OF PATIENT-CENTERED DESIGN}
We took an exploratory approach to answer our research question:
how to design patient health data visualizations. Throughout this process, we gained a rich understanding of the problem, learned valuable lessons for designing future visualizations,
and faced several challenges. In this section, we discuss 
some challenges we encountered on the way.



\textit{(Re-)Recruitment:} Although it is important to understand this 
problem from both perspectives, patients and healthcare providers, 
it is rather challenging to recruit them to take part in research studies. 
Patients
have the extra burden 
of managing their conditions, so it can be difficult for them to allocate 
time for participation in research studies. Recruiting the same participants for future iterations of a design study can be even more challenging as patients may not be physically or mentally stable to participate.
Finding providers willing to give interview time is also a challenge
since they often have a busy schedule or they may be skeptical 
of the value of new technology research. 
This distrust could be due to
their previous experience using ineffective technologies that do not meet 
their expectations.
To overcome these challenges, we collaborated with healthcare providers from a research center (Ward of 21st Century) in Alberta. We worked with several healthcare providers interested and involved in research and through them, we had easier access to recruiting patients. However, when we re-approached patients for a second interview, we could only re-interview one patient.


\textit{Information privacy:} Following ethics board guidelines, we informed participants that we would keep their study data and information private. However, patients are sometimes reluctant to disclose information about their interactions with their healthcare providers since they may have concerns that their healthcare providers 
will have access to this information. To overcome this challenge and create a more open environment, we (interviewers) shared our personal health stories with patients before the interview. On the other hand, healthcare providers also sometimes are hesitant to share their thoughts with researchers. They may have concerns that patients or other colleagues will hear the information.

\textit{Data accessibility:} In some cases, it is helpful to look at patient data to get real-world examples of data. In these circumstances, it is even harder to find participants.
Many patients rely on their memory to keep track of their health data, so they do not have any written record of their data to share with researchers. Also, among those patients who collect, record, and maintain their data, many patients collect data on paper mixed with other personal information.
Thus, researchers need to clean and transform hand-collected data into digital forms.
Patients who use apps or tools to collect  data may not have easy 
access to their data as some of these tools do not
provide an easy way to export or share their data.
To overcome these challenges, we asked patients to walk us through their data, the rationale for their choice of data format and recording device.

\textit{Technology integration:} Electronic healthcare systems are large and complex. Thus, integrating new technologies into these systems can be challenging. The development platforms used for implementing the healthcare systems may not be receptive to incorporating new innovative designs. For instance, the visualization libraries required to develop several designs were not compatible with the patient-centered care plan development platform. Due to limited access to these libraries, the development team needed to spend a significant amount of time developing them from scratch, which was out of budget. As a result, we had to discard several of our visualization designs. 


\section{DISCUSSION}

The results of all the empirical studies we conducted with patients and healthcare providers unfolded the importance of designing for the individual patient while considering their relationship with healthcare providers.
Designing a general software, technology, visualization system that can work for everyone can be appealing and cost-efficient, but 
it will not be an efficient solution for everyone~\cite{Bertelsen2018}. 
Bertelsen et al.~\cite{Bertelsen2018} discuss the benefits of HCI research that focuses on addressing the particular challenges of particular people in particular situations or activities. 
This approach captures the complex and rich nuances of particular people or situations,  which leads to rethinking assumptions and methods we use as researchers. 
Here, we would like to echo the voice of the authors and point to the necessity of designing for particulars, individuals. 
Considering individuality when designing for patients can be even more critical as 
we need to consider biological, psychological, and social factors~\cite{Engel1981}. 


To address our research question, we started with the mindset of designing targeted technological or visualization solutions to represent patient health data. 
Our goal at first was to introduce more point solutions until eventually, we reach a general solution~\cite{Card1997}. Instead, through qualitative research methods and an iterative design process, we came to the conclusion that we need to design individualized solutions by considering each patient. Each patient has a unique body, a highly individualized lifestyle, a different set of goals, and a personalized patient-provider relationship.  
We need to consider all these factors while designing for patients. How can we design only one visualization that can consider all these differences in patients? Can one visualization design fit all?

While we see value in designing for generalization, we first need to start investigating designs for individual patients. Some examples of factors to consider to design individualized visualizations are as follows. When designing patient data visualizations, we need to design by considering patient preference in ways to display missing data; some patients are reluctant to see they have missed collecting data for days, some patients need to have a reminder of missing data to motivate them to collect their data more regularly.
Another factor we found to consider in the design is data context (i.e. food, exercise, life events). Some patients collect contextual data and want to see the relationship between their health data and the contextual data. Other factors to consider to design individualized visualizations are the patient preference in the amount and type of data to share with each healthcare provider, healthcare provider preference in the types of data to review in the clinic, and healthcare provider and patient preference for the mode of data sharing (i.e. hard copy, on the phone, or on a care plan website).


Throughout our studies, we included the perspectives of a small number of patients and healthcare providers. We are aware that we may have not included  other critical perspectives in these studies. Thus, we like to encourage the (human-computer interaction, visualization, healthcare) communities to repeat these studies by including more patients, healthcare providers, healthcare systems and exploring designing visualizations for each individual.
Then, as a community, we need to move towards accumulating these perspectives and designs to empower individuals with accessible design variations.

\section{CONCLUSION}

In this article, we discussed why we think of designing patient health data visualizations as a wicked problem. Drawing from our experience addressing this research question using a patient-centered design approach, we reviewed the lessons we learned in this process. We highlighted the challenges of designing patient health data visualizations in different phases of patient-centered design. These challenges include (re)recruitment, information privacy, data security, data accessibility, technology reception, and technology integration. We concluded with a call to researchers and visualization designers to consider the individuality of patients, healthcare providers, and healthcare systems when designing patient health data visualizations.

\section{ACKNOWLEDGMENT}

We would like to thank our participants for their input. 
This work was supported in part by
the Ward of 21st Century, AITF, and NSERC.






\begin{IEEEbiography}{Fateme Rajabiyazdi}{\,}is
an assistant professor in
the Department of Systems and Computer Engineering
at Carleton University. Before joining
Carleton, she was a postdoctoral researcher at
McGill University Health Center, where she was
a recipient of Fonds de la Recherche en Sant\'{e}
du Quebec postdoc scholarship 2020. She received
her Ph.D. in Computer Science in the
area of information visualization from the University
of Calgary in 2018, and was a recipient of
the W21C health services research scholarship.
Her research interests include visualizations that support patient-healthcare provider communication.
Contact her at fateme.rajabiyazdi@carleton.ca.
\end{IEEEbiography}

\begin{IEEEbiography}{Charles Perin}{\,}
 is an assistant professor in the
Department of Computer Science at the University
of Victoria. He received his Ph.D. from
University Paris-Sud. His research focuses on
designing and studying new interactions for visualizations
and on understanding how people may
make use of and interact with visualizations in
their everyday lives, including mobile and physical
visualization. He has 
served the international community
in multiple roles, such as PC member for IEEE Infovis/VIS
and Graphics Interfaces, and Associate Chair/Editor for ACM CHI and
ACM ISS. He has been in the organizing committee of the IEEE VIS
conference for the past five years and is the Co-General Chair for the
Vis Arts Program at IEEE VIS in 2021.
Contact him at cperin@uvic.ca.
\end{IEEEbiography}

\begin{IEEEbiography}{Lora Oehlberg}{\,} is an associate professor in the
Computer Science department at the University
of Calgary, Canada. She received her doctoral
degree from the University of California at Berkeley,
USA, in 2012. Previously she was a postdoctoral
researcher with the InSitu project team
at Inria-Saclay, France. Her research interests
include human-computer interaction, particularly
in technologies that support creativity, collaboration,
and fabrication.
Contact her at lora.oehlberg@ucalgary.ca.
\end{IEEEbiography}

\begin{IEEEbiography}{Sheelagh Carpendale}{\,} is a full Professor and
Canada Research Chair in Information Visualization
in Computing Science at Simon Fraser
University. Her awards include: IEEE Visualization
Career Award, E.W.R. NSERC STEACIE;
a British Academy of Film \& Television Arts
(BAFTA), an ASTech Innovations in Technology
Award, an NSERC/AITF/SMART Technologies
Industrial Research Chair in Interactive Technologies,
and a CHCCS Achievement Award.
She has been inducted into the both IEEE Visualization
Academy and the ACM CHI Academy. She is an internationally
renowned leader in both information visualization and large display
interaction and has served in roles such as Papers, Program, and
Conference chair for IEEE InfoVis, PacificVis, and ACM ISS and has
received both the IEEE and ACM recognition of service awards. She is
a member of the IEEE Computer Society.
Contact her at sheelagh@sfu.ca.
\end{IEEEbiography}
\balance

\end{document}